# Mining Generalized Graph Patterns based on User Examples[1]

Pavel Dmitriev
*Department of Computer Science*
*Cornell University*
dmitriev@cs.cornell.edu

Carl Lagoze
*Department of Computer Science*
*Cornell University*
lagoze@cs.cornell.edu

**Abstract**

*There has been a lot of recent interest in mining patterns from graphs. Often, the exact structure of the patterns of interest is not known. This happens, for example, when molecular structures are mined to discover fragments useful as features in chemical compound classification task, or when web sites are mined to discover sets of web pages representing logical documents. Such patterns are often generated from a few small subgraphs (cores), according to certain generalization rules (GRs). We call such patterns "generalized patterns"(GPs). While being structurally different, GPs often perform the same function in the network.*

*Previously proposed approaches to mining GPs either assumed that the cores and the GRs are given, or that all interesting GPs are frequent. These are strong assumptions, which often do not hold in practical applications. In this paper, we propose an approach to mining GPs that is free from the above assumptions. Given a small number of GPs selected by the user, our algorithm discovers all GPs similar to the user examples. First, a machine learning-style approach is used to find the cores. Second, generalizations of the cores in the graph are computed to identify GPs. Evaluation on synthetic data, generated using real cores and GRs from biological and web domains, demonstrates effectiveness of our approach.*

## 1. Introduction

Graphs provide a convenient way to represent the structure of data arising in many chemical, biological, social, technological, and web applications. Hence, there has been increasing demand for automatic methods for mining useful information from graph data. Typically, such information is represented in the form of *graph patterns* – subgraphs having specific structure [19]. For example, different graph patterns have been used to cluster protein interaction and gene co-expression networks [6, 16], and to discover communities on the Web [11]. When the exact structure of the patterns of interest is not known, subgraphs occurring frequently in the graph were often used. Applications include, among others, characterizing behavior of molecules [7], classifying chemical compounds [2, 12], and discovering patterns in semistructured web data [15].

While frequent subgraphs may serve as good attributes for describing graphs in applications such as graph clustering or classification, often one is interested in mining patterns that perform the same or similar functions in the network. Being able to identify such patterns would allow for more accurate frequent subgraph mining, and has many applications on its own [14]. For example, in the biological domain, one may be interested in studying evolution of "functional modules" (groups of interacting molecules performing a specific function in the network) [16]. Such modules arise through duplication of useful fragments and their evolution over time. They may have slightly different structure and do not necessarily occur frequently in the networks of interest. In the web domain, one might be interested in identifying "logical documents" (sets of web pages representing semantically coherent documents) on web sites [3]. A typical example of a logical document is an article on the web, physically consisting of the contents page, and several HTML pages with content (Fig. 1a). In this case, the task is to find all sets of pages on the web site corresponding to web articles. As in the previous case, articles may have different number of pages, not all of which are frequent. Fig 1b shows a fragment of a web site with two articles on it.

---
[1] A short version of this paper appeared in the proceedings of ICDM-2006 conference.

The above examples demonstrate the need for a different approach to graph mining, and it is this problem that we focus on in this paper. We assume, as in the previous work [9, 14], that the patterns of interest are generated from a set of basic elements, or *cores* – small subgraphs of specific structure, according to some *generalization rules* (GRs). We call such patterns *generalized patterns* (GPs), and the problem of finding GPs *generalized pattern mining*. An example of such generalization procedure for the case of a web article is given in Fig. 1c.

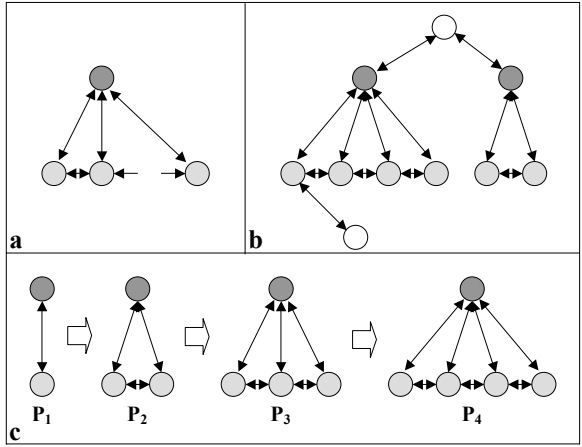

**Figure 1. (a) A typical web article pattern;
(b) Two occurrences of the pattern in a web graph;
(c) Generalizations of a web article pattern.**

One way to approach the above problem is to fix the set of cores and generalization rules. Solutions to specific instances of such approach have been presented in [9, 14, 7]. However, this requires good understanding of the network structure and the GPs of interest, it is network and task-specific, and incurs considerable human effort to specify cores and GRs.

In this paper, we propose a new approach to GP mining. The approach consists of the following steps.
- A user provides a small number of examples of GPs he or she is interested in;
- Cores are generated using the examples specified by the user;
- GPs are extracted by generalizing the cores according to GRs created automatically based on the structure of the cores.

Note that our approach makes no apriori assumptions about the structure of the cores, or frequency of GPs in the graph, and only requires the user to proved a few examples of GPs of interest.

To simplify explanation, we only consider the case of directed labeled graphs in this paper. It is straightforward to generalize the algorithms to the case of undirected and/or unlabeled graphs.

## 2. Preliminaries

A *directed graph g* consists of a set of vertices, $V(g)$, and a set of edges, $E(g)$, where an edge $e \in E(g)$ is an ordered pair $(v_1, v_2)$, $v_i \in V(g)$. A *labeled graph* has labels associated with its vertices and edges. *Label function l* maps a vertex or an edge of a graph to its label. A graph $g$ is a *subgraph* of graph $g'$ if there exists a subgraph isomorphism from $g$ to $g'$.

A *subgraph isomorphism* between $g$ and $g'$ is an injective function $f : V(g) \rightarrow V(g')$, such that (1) $\forall v \in V(g)$, $l(v) = l'(f(v))$, and (2) $\forall$ edge $(u, v) \in E(g)$, $(f(u), f(v)) \in E(g')$ and $l((u, v)) = l'((f(u), f(v)))$, where $l$ and $l'$ are label functions of $g$ and $g'$ respectively. Sometimes, if $g$ is a subgraph of $g'$, we will say that $g$ matches $g'$. A mapping between vertices and edges of $g$ and $g'$ is also called an *embedding* of $g$ in $g'$.

Let $G$ be the set of all directed labeled graphs, and let $\bar{g}$ denote $\{g'\} \subset G$, such that for every $g'$, $g$ is a subgraph of $g'$, and $|V(g)| < |V(g')|$.

**Def. 1 (generalization rule).** A *generalization rule* (GR) is a function $r : G \rightarrow 2^{\bar{g}}$. If, $\forall g, r(g)$ consists of graphs $g'$ s.t. $|V(g)| = |V(g')| - 1$, the generalization rule is called *simple*.

Let $g$ be a graph, and $c$ be a subgraph of $g$. Graph $g'$ is an *extension of c in g by r*, or $g' = \psi_r(c)$, if $g'$ is a subgraph of $g$, and $g' \in r(c)$.

**Def. 2 (generalized pattern).** Graph $g'$ is a *generalized pattern* (GP) of $g$ induced from the core $c$ by GR $r$, if (1) $g'$ is an induced subgraph of $g$, (2) $\psi_r(g') = \varnothing$, (3) $g' = \psi_r^n(c)$ for some $n$.

The task of generalized pattern mining is, given a graph $g$, a GR $r$, and a core $c$, to find all GPs induced from $c$ in $g$ by $r$.

GP mining with arbitrary $g$, $c$, and $r$ is a hard problem. It is not difficult to show that the problem is NP-hard (by reduction from subgraph isomorphism problem). However, we were not able to show whether it is in NP class. While high complexity is an issue, in practice the problem can often be solved efficiently, since the GPs of interest are typically small, and one can make use of the properties of the GR under consideration. A bigger problem is that specifying cores and GRs puts a huge burden on the user. It is this problem that we focus on in this paper.

## 3. Related work

Most of the previous work in pattern mining in graphs considered static patters [10, 20, 18]. An algorithm for mining molecular fragments using wildcards was presented in [7]. A wildcard allows a pattern vertex to match vertices with different labels in the graph, thus allowing the pattern to have non-isomorphic embeddings. In [14], an algorithm is described that allows for discovery of frequent molecular fragments containing chains of atoms of varying length. Patterns that differ only in the length of a chain are considered identical for support computation purposes. While in [14] the only kind of GPs considered are chains of atoms of a particular type, the frequent fragment mining algorithm they present seems to be generalizable to other types of GPs. However, both [7] and [14] rely heavily on the fact that the wildcards and the chains are provided by the user.

A more relevant work is [9], where the authors studied two general types of GRs applicable to any subgraph. First, every node in the subgraph is assigned a "role". Two nodes have the same role if there exists an automorphism that maps them into each other, and have different roles otherwise. Then, generalization of a subgraph is performed by duplicating a role and its connections. In [9], such kinds of generalizations have been observed in biological and technological networks.

However, there are problems with applying such generalization in practical applications. For example, it is not possible to generalize from $P_3$ to $P_4$ for the web article example on Fig. 1c. On the other hand, if pattern $P_1$ from Fig. 1c is taken as a core, its weak generalization applied to the graph on Fig. 1b will return the whole graph. Finally, similar to the previous works, [9] assumes that the cores are provided by the user.

In our paper, we address the limitations of the works mentioned above. We address the problem of identifying cores by proposing a method for automatically extracting cores from the graph based on user examples. We also propose a new way to generate a simple GR based on the structure of the core. We analyze when exactly our rule may produce subgraphs that are too general. We show that it depends on the cores used, as well as on the structure of neighborhoods of GPs, and prove that, given "good" user examples, our GR never produces too general GPs, when used with the cores generated by our algorithm.

## 4. Mining generalized patterns from user examples

In this section, we describe our algorithm for mining GPs from user examples. Given a graph g and a set of example GPs specified by the user, our algorithm finds all GPs in the graph that are similar to the user examples.

The algorithm consists of the following steps:
- Extracting negative examples;
- Generating GP cores;
- Using the cores to identify GPs in the graph.

There are several assumptions that we make in order to ensure the correctness of the algorithm. First, we assume that the user examples are indeed generated from some set of cores using some GRs, which are not known to the user. Second, we assume that GPs in the graph do not overlap.

While the latter assumption is essential for correctness of negative example generation, and for the theoretical guarantees proved below, the algorithm can be modified to include the case of overlapping GPs. However, this requires that the set of user examples either does not contain overlapping GPs, or, if it contains a GP overlapping with some other GP in the graph, it also contains all GPs it overlaps with. Similar, but slightly weaker theoretical guarantees hold in this case as well. However, since the proofs and the descriptions of the algorithms are significantly more complicated in this case, we do not describe it in the paper.

In the following sections, we discuss each of the steps of our algorithm in detail.

### 4.1. Extracting negative examples

Intuitively, negative examples are subgraphs of g which are not GPs. They play an important role in the process of generating GP cores. Namely, they keep cores from becoming too general (see section 4.2).

---

Input: graph $g$, set of positive examples $P^+ = \{p_1^+,...,p_n^+\}$, max. size of a negative example $k$, set of negative edges $N_e = \bigcup_1^n N_e(p_i^+)$.
Output: set of negative examples $S$.
    Algorithm:
- $S$ = set of all subgraphs of size at most $k$, containing at least one edge from $N_e$
- For every $p$ from $S$ do
    o For every $p^+$ from $P^+$ do
        ▪ If $p^+$ matches $p$ (not including negative labels) do
            • $S = S \backslash \{p\}$
- For every pair of $p_1, p_2$ from $S$ do
    o If $p_1$ matches $p_2$ (including negative labels) do
        ▪ $S = S \backslash \{p_1\}$

**Figure 3. Algorithm for extracting negative examples.**

Consider a graph *g*. Let $p^+$ be an example GP specified by the user. We define an *edge neighborhood* of $p^+$, $N_e(p^+)$, to be a subset of edges of *g* having one end in $V(p^+)$, and the other end in $V(g)\backslash V(p^+)$. Similarly, a vertex neighbourhood of $p^+$, $N_v(p^+)$, is $V(N_e(p^+))\backslash V(p^+)$. Since user examples do not overlap, any subgraph of *g* containing at least one edge in the edge neighbourhood of a positive example cannot be (a part of ) a GP. Thus, we can take as negative examples all non-isomorphic subgraphs of *g* up to specified size that contain a negative edge, and are not matched by a positive example. We assign a special "negative" label to the edges from $N_e(p^+)$, which we call *negative edges*. These labels are used in the graph isomorphism computation to select non-isomorphic negative examples. It is also possible that there will be two negative examples such that one is a subgraph of the other. In this case, we keep only the larger example. The algorithm is presented in Fig. 3. Fig. 5b shows negative examples generated for the graph on Fig. 5a. Positive examples are circled, and negative edges are highlighted in grey.

We would like to note that, for the algorithm in Fig. 3, as well as for other algorithms presented in this paper, we only show the most straightforward implementation. The actual implementation often uses auxiliary data structures and different order of operations to speed up the computation. Due to the lack of space, and to simplify the explanation, we omit the discussion of these details.

### 4.2. Generating GP cores

To generate GP cores, we use an algorithm similar to the one described in [8] (see section 3). Starting from positive examples, the algorithm builds a lattice by recursively computing sets of maximal common subgraphs of pairs of positive examples. In [8], only the subgraphs that do not match any negative example are included in the next level of the lattice. Here, we use the concept of *strong matching* instead.

**Def. 3 (strong matching).** Given a negative example $p^-$, and a subgraph $g'$, we say that $g'$ *strongly matches* $p^-$ if $\exists$ at least one embedding of $g'$ in $p^-$ that contains a negative edge.

We include in the next level of the lattice only the subgraphs that do not strongly match any negative example. When the computation stops, the subgraphs highest in the lattice (positive hypotheses) are taken as cores.

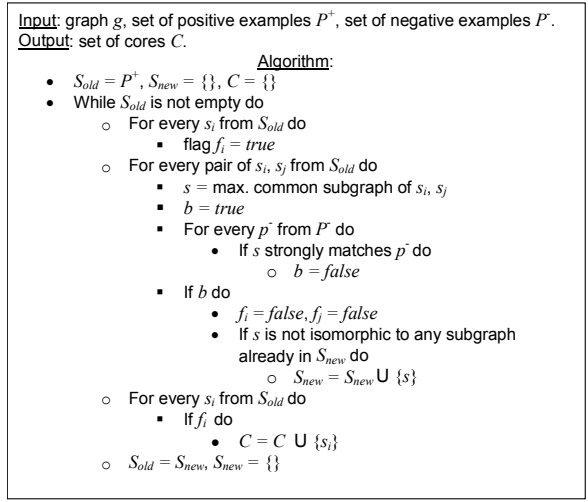

Input: graph *g*, set of positive examples $P^+$, set of negative examples $P^-$.
Output: set of cores *C*.
Algorithm:
- $S_{old} = P^+$, $S_{new} = \{\}$, $C = \{\}$
- While $S_{old}$ is not empty do
  - For every $s_i$ from $S_{old}$ do
    - flag $f_i = true$
  - For every pair of $s_i$, $s_j$ from $S_{old}$ do
    - $s$ = max. common subgraph of $s_i$, $s_j$
    - $b = true$
    - For every $p^-$ from $P^-$ do
      - If *s* strongly matches $p^-$ do
        - $b = false$
    - If *b* do
      - $f_i = false, f_j = false$
      - If *s* is not isomorphic to any subgraph already in $S_{new}$ do
        - $S_{new} = S_{new} \cup \{s\}$
  - For every $s_i$ from $S_{old}$ do
    - If $f_i$ do
      - $C = C \cup \{s_i\}$
  - $S_{old} = S_{new}$, $S_{new} = \{\}$

**Figure 4. Algorithm for generating GP cores.**

The intuition behind using strong matching is that it is possible that a subgraph of a negative example is a part of a GP, if it does not include any negative edges. Thus, in order to conclude that an element of the lattice is not a core, it has to match at least one negative edge in a negative example.

The core generation algorithm is presented in Fig. 4. Fig. 5c shows the lattice built from the positive examples on Fig. 5a, and negative examples on Fig. 5b, with the resulting cores circled.

However, during our experiments, we found that the cores output by the algorithm on Fig. 4 are often too large. For example, when experimenting with the web article pattern from Fig. 1, we found that subgraph $P_3$ is occasionally output as a core. This subgraph consists of 4 vertices, which prevents us from discovering GPs consisting of 2 or 3 vertices. In general, in practical applications, GRs are sometimes such that there is a unique (up to isomorphism) way to extend a core to an arbitrary size pattern. In this case, the set of cores (positive hypotheses) output by our algorithm consists of a single core isomorphic to the smallest positive example, preventing us from discovering smaller GPs. To deal with this problem we introduce a new concept, *hypothesis relaxation*.

**Def. 4 (hypothesis relaxation).** Given a positive hypothesis *h* and a set of negative examples $P^-$, a *relaxation* of *h* is a set $\{h'\}$ of subgraphs of *h*, s.t. (1) neither of $h'$ strongly matches any $p^- \in P^-$, and (2) $\forall h', \forall h''$ subgraph of $h'$, $h''$ strongly matches at least one $p^- \in P^-$.

We apply hypothesis relaxation to every core output by the algorithm on Fig. 4, and we take as the final set of cores the set of all non-isomorphic subgraphs found by the relaxation procedure. Due to the lack of space, we omit the formal description of the hypothesis relaxation algorithm. Cores resulting from relaxation of the hypotheses from Fig. 5c are shown on Fig. 5d.

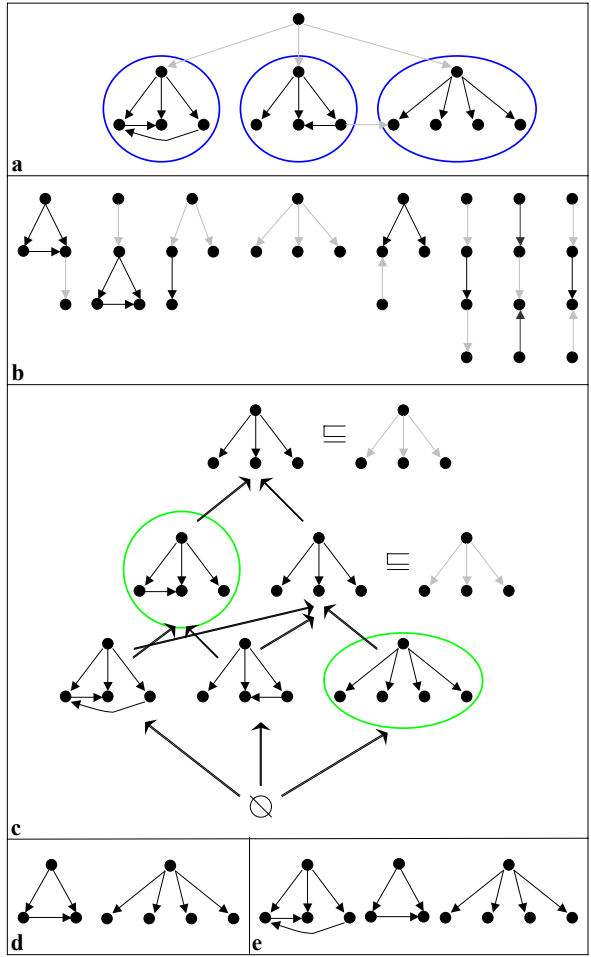

**Figure 5. (a) Example graph; user-specified (positive) examples are circled, negative edges are highlighted in grey;
(b) Negative examples;
(c) Positive lattice generated by our machine learning algorithm; positive hypotheses are circled;
(d) Positive hypotheses after relaxation;
(e) Subgraphs found by the GP mining algorithm.**

## 4.3. Mining generalized patterns

In this section, we describe the GP mining algorithm. Given a graph $g$, a core $c$, and a GR $r$ the algorithm finds all occurrences of the GPs induced from $c$ in $g$ by $r$.

Before describing the algorithm, we need to define the GR we will use. We define a *natural expansion rule* as follows.

**Def. 5 (natural expansion rule).** Given a core $c$, $|V(c)| = m$, the *natural expansion rule* $r_c$ is a simple GR s.t. $\forall \ g' \in r_c(g)$, if $v \in V(g')\setminus V(g)$, then there exist $v_1,\ldots,v_{m-1}$, $v_i \in V(g')$, s.t. $c$ matches the induced subgraph of $g'$ defined by $\{v_1,\ldots,v_{m-1},v\}$.

Essentially, natural expansion rule works by adding a copy of $c$ to $g$ in such a way that $m-1$ vertices are taken from $g$, and only one new vertex is introduced. For example, this rule can be used to perform the generalizations on Fig. 1c.

Natural expansion rule possesses a nice property when applied to GP mining.

**Proposition 1.** *Given an arbitrary graph $g$, a core $c$, and an embedding $s$ of $c$ in $g$, there exists a unique GP $p$ induced from $c$ in $g$ by $r_c$, containing $s$.*

It follows from proposition 1 that, given an embedding of a core in the graph, it does not matter in which order we expand the embedding to obtain a GP, since the GP is unique. This property is used in our GP mining algorithm, presented on Fig. 6. Note also that proposition 1 does not hold for an arbitrary GR. For example, it does not hold for either strong or weak GR described in section 3.

As we mentioned in section 3, a problem with the natural expansion rule is that it may produce too general GPs. We will now prove that, under reasonable conditions, the algorithm on Fig. 6 can never produce too general results, when applied to the cores generated by the algorithm on Fig. 4.

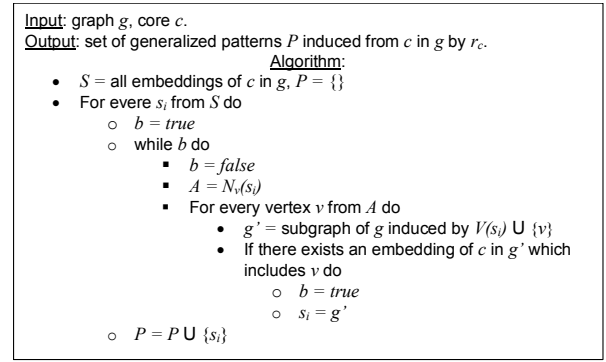

**Figure 6. Algorithm for generating GPs using natural expansion rule.**

Let $g$ be a graph, $P^+$ be a set of user-provided GPs, and $P$ be the set of all true GPs in $g$. Let $E^-_{P^+}$ be the set of negative edges obtained from $P^+$, and $E^-_P$ be the set of true negative edges (all edges which are not within any GP). Let $Neg_{P^+}$ be the set of negative examples generated starting from $E^-_{P^+}$, and $Neg_P$ be the set of negative examples generated starting from $E^-_P$ by the algorithm on Fig. 4. Finally, let the maximum size of a negative example be $k$.

**Theorem 1.** *Let C be the set of cores generated by the algorithm on Fig. 4 from $Neg_{P^+}$, and let S be the set of GPs generated from C by the algorithm on Fig. 6. If (1) $Neg_{P^+} = Neg_P$ (up to isomorphism), (2) no user example matches any $p^- \in Neg_{P^+}$, and (3) $\forall c \in C, |V(c)| \le k$, then $\forall s \in S, \exists p \in P$ such that s is a subgraph of p.*

**Proof.** Suppose that there exists $s$ in $S$ such that $s$ is not a subgraph of any $p$ in $P$. Let $c$ be a core that generated $s$. By (2), there must exist an edge $e^- \in E(s)$, such that $e^- \in E^-_P$. Let $s'$ be the subgraph isomorphic to $c$ used to include the last end of $e^-$ in $s$, according to the algorithm on Fig. 6. There are two cases. <u>Case 1</u>: $s'$ contains $e^-$. Since, from (3), $|V(s')| \le k$, there exists a negative example in $Neg_P$, and $s'$ strongly matches it. From (1), it follows that there exists a negative example $q^-$ in $Neg_{P^+}$ such that $s'$ strongly matches it. Since $s'$ is isomorphic to $c$, $c$ strongly matches $q^-$. But this means that $c$ could not be produced by the algorithm on Fig. 4. Contradiction. <u>Case 2</u>: $s'$ does not contain $e^-$. Then, if both ends of $e^-$ are in $s'$, there exists a negative example in $Neg_P$ containing $e^-$, and we get to a contradiction by the argument similar to case 1. Otherwise, there must exist some other negative edge, $e^=$, sharing a vertex with $e^-$, that is in $s'$. Again, an argument similar to case 1, applied to $e^=$, leads to a contradiction. □

Essentially, theorem 1 says that, if the problem is well-defined (condition (2)), then, given "good" user examples (conditions (1) and (3)), our algorithm will never produce GPs that are too general.

### 4.4. Complexity of the algorithm

In this section, we analyze the complexity of our algorithm. As we noted earlier, by using auxiliary data structures and changing order of operations in some algorithms, algorithms in Fig. 3, Fig. 4, and Fig. 6 can be improved. Since optimization of the algorithms is not the focus of this paper, we do not present analysis of these improved versions of the algorithms here. However, we believe that GP mining problem has a great potential for the use of approximation algorithms and heuristics to reduce the search space on different steps of the process. This is one of the topics of our future work.

We assume that the maximum size of a negative example, and the number of user-specified examples are constant. We denote maximum size of a negative example by $k$, maximum size of a user-provided example by $b$, the number of negative examples by $m$, and the set of cores generated by our algorithm by $C$.

To extract negative examples, we perform $O(|E(g)|)$ operations to extract negative edges, $O(|E(g)|)$ subgraph isomorphism computations to select the negative examples, and $O(|E(g)|)$ subgraph isomorphism computations to eliminate subgraphs matched by a positive example. Note that the subgraph isomorphism computations are performed on subgraphs of size at most $max(k, b)$.

To generate cores, we perform $O(b)$ maximum common subgraphs computations, $O(b)$ graph isomorphism computations (to eliminate isomorphic elements), and $O(b*m)$ subgraph isomorphism computations (to eliminate elements matching a negative example, and to perform hypothesis relaxation). Again, the size of the subgraphs for which the computations are performed is at most $max(k, b)$.

Finally, to generate GPs, we perform $O(|V(g)|*|C|/min_c(|V(c)|))$ subgraph isomorphism computations to find embeddings of the cores[2], and $O(|V(g)|)$ subgraph isomorphism computations to expand one embedding, total of $O(|V(g)|^2*|C|/min_c(|V(c)|))$ computations. The latter computations are performed on subgraphs of size at most $max_c(|V(c)|)$.

As one can see, the complexity of the algorithm is rather high, requiring a polynomial number of subgraph isomorphism, graph isomorphism, and maximal common subgraph computations. Exact algorithms for these problems are known to have exponential running time in the worst case. However, almost all of these computations are performed on subgraphs of very small size, bound by the $max(k, b, max_c(|V(c)|)$. In the practical applications we are aware of, this number is not likely to exceed 10. Thus, we believe that the algorithm is practical to apply to real life applications.

## 5. Experimental results

In this section, we present an experimental evaluation of our algorithms. Experiments were conducted on synthetic data generated using cores and GRs frequently encountered in biological and web domains.

---
[2] This is achieved by finding embeddings one at a time, computing a GP from this embedding, and deleting the GP from the graph before the next embedding is computed.

## 5.1. Experimental Setup

We experimented with four cores and GRs: $BP_1$ and $BP_2$ from the biological domain, and $WP_1$ and $WP_2$ from the web domain. Fig. 7 shows the cores and illustrates the generalization procedure. $WP_1$ is the web document pattern we discussed earlier (see section 1). $WP_2$ is another common web document pattern, which has much simpler structure than $WP_1$. $BP_1$ is the feed-forward loop pattern mentioned in section 3, and $BP_2$ is another common biological pattern called bi-fan. Generalizations of $WP_1$ and $WP_2$ were observed to occur frequently on web sites in our earlier experiments with logical documents [3]. Strong and weak generalizations of $BP_1$ and $BP_2$ were reported to often occur in biological and technological networks by [9].

For each of our experiments, we generated 20 GPs by choosing the number of vertices in a GP, as well as the roles for the biological GPs, at random. We then connected the GPs into a graph by introducing links between them. The source and the target of a link were chosen at random so that they do not belong to the same GP. Thus, there were no random links within the GPs.

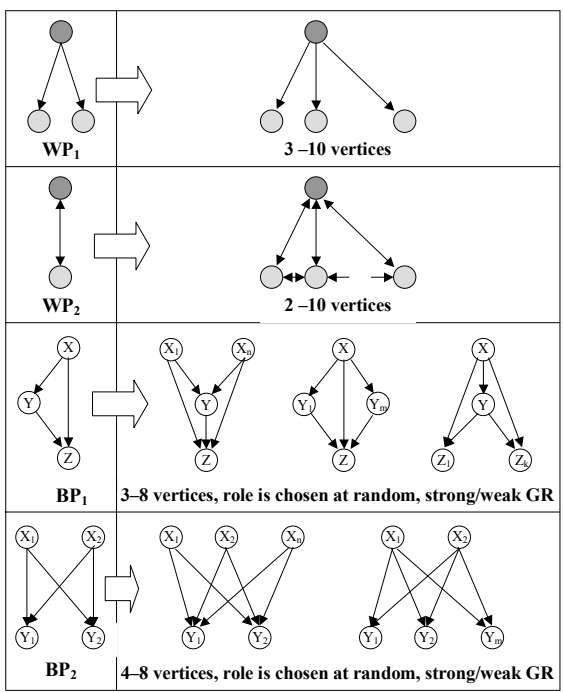

**Figure 7. Cores and GRs used in our experiments.**

Using recall and precision as performance measures, we studied the effect of the number of random links introduced, and the number of user examples on performance. Every reported value is an average over 10 runs of the experiment. In our preliminary experiments we did not see any significant difference in performance depending on the maximum example size for generating negative examples (we tried sizes 4 and 5). Thus, the value 4 was used for all experiments described below.

## 5.2. Results

First, we evaluated our algorithm in a setting when 3 of the GPs were chosen at random as user examples, and between 10 and 200 random links were added among the GPs to form a graph. The results are shown in Fig. 8.

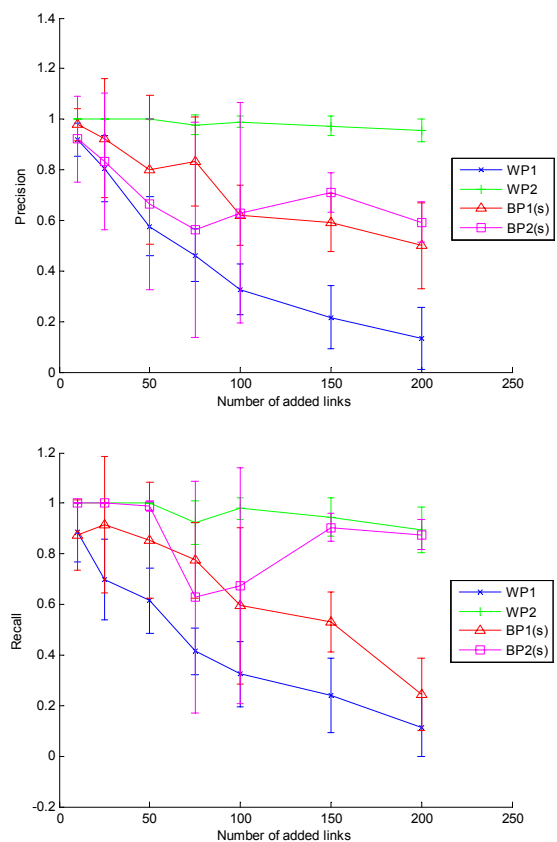

**Figure 8. Performance for different number of added random links, using 3 user examples. Strong GR was used to generate BP1 and BP2.**

Fig. 8 shows that, when the number of added random links does not exceed 50, the algorithm performs well for all cores. However, after that performance differs for different patterns. $WP_2$, which has the structure that is difficult to obtain with random links, shows very good performance even when 200 random links are added. $WP_1$, on the other hand has a very simple structure that can easily appear at random. Not surprisingly, its performance decreases sharply as more random links are added. Biological patterns, $BP_1$ and $BP_2$, show performance in between $WP_1$ and $WP_2$. They show similar precision, but $BP_2$ has significantly higher recall. We believe the difference in recall is due to $BP_2$ being denser than $BP_1$, and, therefore, harder to be reproduced by random links.

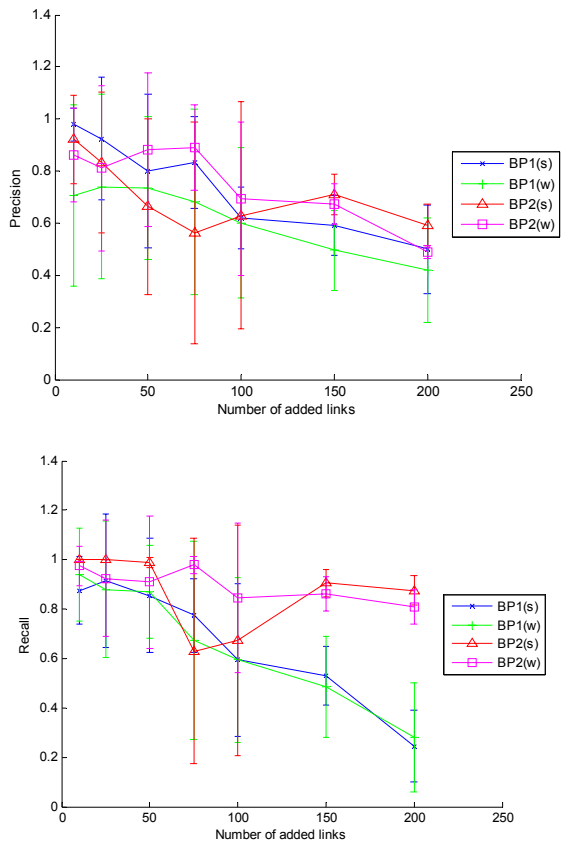

**Figure 9. Comparison of performance for BP1 and BP2 patterns, when strong (s) and weak (w) GRs were used.**

One can also see that, for the biological patterns, the standard deviations are very high when the number of added random links is between 25 and 100, in the middle of the range, and they are smaller at the ends of the range. This is because in the middle of the range the number of potential non-isomorphic negative examples that may arise with reasonable probability is higher. Thus, different sets of negative examples appear at every run of the algorithm, which leads to discovery of different cores, and generation of different GPs. The web patterns do not suffer from this effect, because the number of potential cores for them is very limited due to their simple structure.

A similar effect can be observed on Fig. 9, where we compare the results for strong and weak GRs applied to the biological patterns. Again, standard deviations are higher in the middle region.

Fig. 9 shows that, overall, the results for the biological patterns using strong and weak GRs are similar, with weak GRs having slightly lower performance, and slightly higher standard deviation in most cases. In subsequent experiments, we only use strong GRs for the biological patterns.

Next, we evaluate the impact of the number of user examples on performance. We run our experiments on graphs created with 50 added random links, varying the number of user examples from 1 to 5. Fig. 10 shows the results of this experiment.

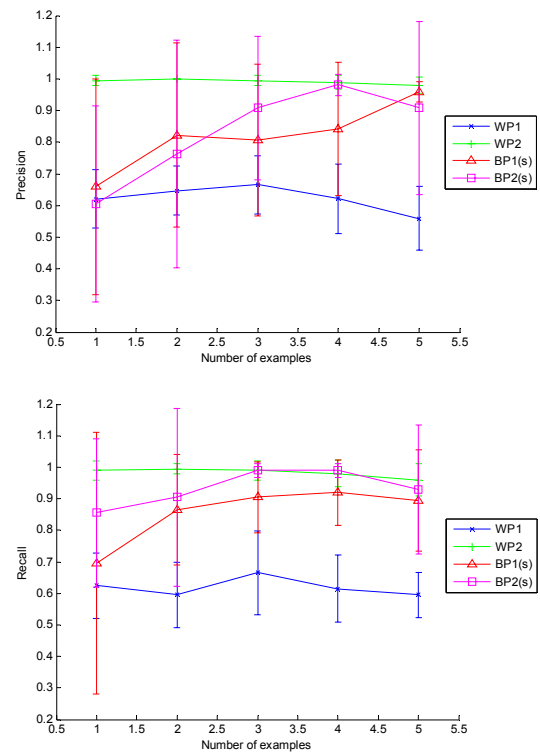

**Figure 10. Performance for different numbers of user examples, using 50 added random links.**

Interestingly, the number of user examples does not have any significant impact on the web patterns. However, this is not the case with the biological patterns – both precision and recall benefit from having at least 3 user examples. We suspect that this might be due to different generalization kinds of rules used to form web and biological GPs; however, we do not know what properties of the rules caused this effect. Getting a better understanding of the kinds of GRs used in practice, and their effect on the performance of our algorithm is one of the directions for our future work.

Finally, we experimented with using only a fraction of all negative examples, as an approximation to the complete set. Since our implementation generates negative examples by repeatedly expanding existing examples through adding a new edge, we can take only a part of the incomplete examples on every step, throwing away the others. By doing that, we hope to improve the running time of the algorithm, since generating negative examples, especially of size 5 or higher, accounts for a significant portion of it. Fig. 11 shows the results for the experiment, when only 10% of the incomplete examples were used on every iteration.

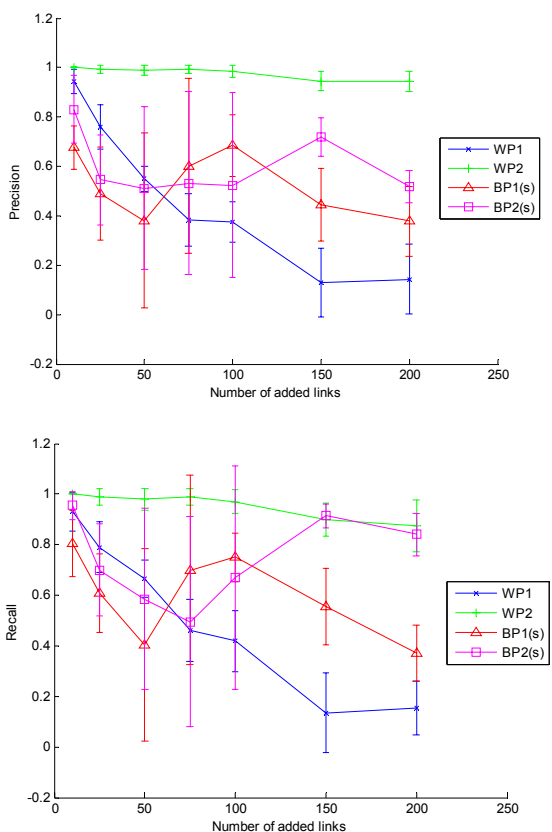

**Figure 11. Performance for different numbers of added random links, using 3 user examples, and 10% of the incomplete examples on every iteration.**

Comparing these results to Fig. 8, one can notice several interesting trends. Similar to the experiments on Fig. 10, performance for the web patterns is not affected by the approximation. For the biological patterns, however, performance drops in the middle of the range, when 25 to 75 links are added. For all patterns the standard deviations are higher than in the results in Fig. 8. These results suggest that, while using an approximation to the complete set of negative examples improves the running time, it may decrease the performance in some cases. More research is needed to understand which properties of patterns make it appropriate to use this approximation scheme. Also, more sophisticated approaches to decide which examples to include may produce better results.

Finally, we would like to say a few words about the running time of our algorithm. For all patterns, an experiment finished in less than a second most of the time. However, some particular runs could take minutes or even hours. From manual examination, we found that this happens in settings with high number of added random links, when the real core is subsumed by a negative example. This results in many larger cores being generated. Such cores are expensive to generate, and they also lead to increased running time of the GP mining algorithm using them. It happened particularly often with biological patterns, and caused us to use smaller value for the maximum number of vertices in experiments with $BP_1$ and $BP_2$. However, we believe that optimization techniques can be used to significantly improve the running time even in these cases. This is one of the directions for our future work. Since the variance in the running time is extremely high, we do not report any detailed results here.

Overall, the experiments show that our GP mining algorithm performs well for patterns from both the web and the biological domains. We expect it to perform even better in practical applications, both in terms of accuracy and in terms of efficiency. First, due to less randomness, there likely to be less non-isomorphic negative examples in practice. Second, a vertex and/or edge labeling of the graph will help to improve the performance. Applying our algorithm to real-life problems is another part of our future work.

## 6. Conclusion

In this paper, we presented a new algorithm for the GP mining problem, which asks, given a graph, a set of cores, and generalization rules, to find all GPs in the graph induced from the cores by the GRs. This problem has important applications in the biological, web, and other domains, and may be used to improve the accuracy of frequent subgraph mining. A problem with applying GP mining in practice, however, is that often neither cores, nor GRs are known to the user. The algorithm we presented in this paper performs GP mining based only on a few examples of GPs selected by the user. In fact, as our experiments showed, sometimes just a single example is enough for our algorithm to produce good results. We proved that, given some reasonalble assumptions, all GPs found by our algorithm would be either equal to, or contained as subgraphs in the real GPs. Experiments on synthetic datasets using real cores and GRs from the biological and web domains showed that our algorithm still performs well even when these assumptions do not hold.

This work can be extended in several directions. First, there is a potential to improve the running time of the algorithm through approximation and optimization techniques. However, as our experiments demonstrated, better understanding of the properties of cores and GRs is needed to determine when approximations are appropriate. Second, there is also a potential to apply active learning to improve the core generation procedure, in particular in cases when the assumptions of theorem 1 are violated. Additional information from the user may help to generate new cores and eliminate the incorrect ones. Finally, we are looking forward to applying our algorithm to real-life problems.

## 7. Acknowledgements

The authors would like to thank Paul Ginsparg for pointing out references [1] and [9], and Thorsten Joachims for valuable discussions. This work is supported by the National Science Foundation Division of Information and Intelligent Systems (IIS) through grant number IIS-0430906 (September 2004 through August 2007).